# The effect of pack formation at the 2005 world orienteering championships

Prof. Graeme J. Ackland
School of Physics, University of Edinburgh
Edinburgh
EH9 3JZ

**Abstract**

At the 2005 world championships there was considerable discussion and a formal protest in the long distance race arising from a perceived advantage obtained by some athletes running together.  It is shown that a statistical model presented in previous work [1] is applicable to this event, giving predictions of the final times to with 2-3 minutes.  Using the model, we show that pack formation was inevitable in this format. The statistical benefit gained at the elite level from running with other competitors appears to derive both from increase speed through the terrain, and the elimination of large navigational errors.

**Introduction**

The rules of orienteering forbid competitors in individual competitions from following one another, however there is no prescription for what should happen when one competitor catches another.  This contrasts with sports such as triathlon where "drafting" (following closely behind another cyclist to obtain reduced aerodynamic resistance) may be forbidden by requiring caught competitors to drop back.  There is, however, little doubt that having sight of another orienteer can affect performance, whether through increased chance of spotting a control marker, improved navigation, reduced need to read the map, finding of better routes through terrain, or simply motivation to run faster.

IOF rule 26.1 states,  "In an individual interval start race, competitors are expected to navigate and run through the terrain independently."  In practice, violations of this rule are commonplace.

The purpose of this paper is to estimate the statistical benefit derived by elite athletes from running together in the recent World Championships.  The study is based on a model for pack formation, described below, which has been tested previously on data from public races in the UK.

**Methods**

In the model, an orienteering race is simulated by assuming that each competitor *i* procedes around the course at a speed $u_i$ determined by their own ability.  When another competitor is within a certain range *r* ahead, this speed is increased by a factor *b*, which is referred to as the boost – these quantities are the same for all competitors.  A competitor obtains no advantage from a following competitor.  We integrate equations of motion in discrete timesteps for interacting competitors moving in one dimension from start to finish.  A faster competitor (or pack) can pass a slower one if, despite the effect of the boost, the slower competitor's speed remains below that of the faster competitor.  There is no provision the model for one competitor to pass another without seeing each other, e.g. if they are taking different routes.  This has some bearing on our results, and will be discussed later.

In previous study, based on a large number of statistically generated artificial races, it was shown that groups would dominate a competition if the fraction of competitors of able to catch another competitor but not get away from them exceeded 13%.  Above this threshold the packs will continue to grow in size.  Insufficient data from actual events was available for detailed statistical analysis (largely because of lack of start time data), but comparison suggested the model was accurate in the absence of large navigational mistakes.

For application to a specific race, it is necessary to determine the individual speeds $u_i$ and the boost factor $b$. Because of the correlation with start time, $u_i$ cannot be taken from the qualifying races. The World Rankings list cannot be used as it is compiled in a way that makes it impossible to relate points to relative running speeds, and in any case conflates results from many different terrain types over a long period of time, which may not be relevant to ability in Japanese terrain in August 2005. Consequently, it was decided to use the average speed over the first third of the course, subtracting errors [2] of more than 45s to derive $u_i$. This has the advantage of being in appropriate terrain and time, and is relatively unaffected by pack formation. By analogy with previous UK work, the boost was set to 8%. Tests with lower boost factors gave poorer fits to the results. With this level of boost, a pair of athletes will move 4% faster than their average, unboosted speed.

The information regarding start times and intermediate lap times was taken from the official website of the 2005 World Championships [2]. These times are not repeated here, but it may be a convenience to the reader to take a copy.

**Results**

Both men's and women's races were examined, comparing the simulated race with the   Good agreement is obtained between the two sets, with packs forming and breaking in the same way for both. It is not the intention of this paper to interfere with the protest in the men's race, so we concentrate our analysis on the women.

In particular, the crucial features of the women's winners are well reproduced: and a discussion of this illustrates most of the salient features of the model. Firstly Jukkola catches 2 mins. on Novikova. In the model these two form a stable pack, however shortly afterwards they are caught by Niggli. Niggli/Jukkola form a stable pack (NJ), but this pack moves too fast for Novikova, who is then dropped. Next two further pairs (Allston/Haapakoski AH and Ryabkina/Jurenkova RJ) are caught by NJ – at one point six of the fastest eight competitors appear to be together – although these groups meet slightly later in the simulation than in the actual race, the model correctly predicts that only Haapakoski from these pairs can stay with NJ, and the boost they provide for her means that Allston will be dropped to finish alone. As can be seen in figure 1, other features of the race are similarly well reproduced, as are details of the men's race.

It is interesting to examine whether effects of packing in the model presented here give it significant additional predictive power over simple extrapolation. In figure 2 the actual results of the race are compared to the model predictions. There is a four minute gap in the results at around 95 minutes, and it can be seen that the model based on the first six controls predicts the finishing times of this group to within three minutes. The agreement with the second group is weaker, because they are more prone to unpredictable navigational errors. Six minutes is the largest deviation. This should be contrasted with naïve extrapolation from the first six controls, which gives errors about 250% greater. It is thus clear that the collective packing behaviour incorporated in the model is important in determining the results.

In the simulated race, the amount of time boosted by being in packs exceeded 40mins for six competitors, five from the leading group (recall that in a pair, only one competitor is boosted at any given time). For a competitor taking 80mins, this level of boost gives a time advantage of around three minutes.

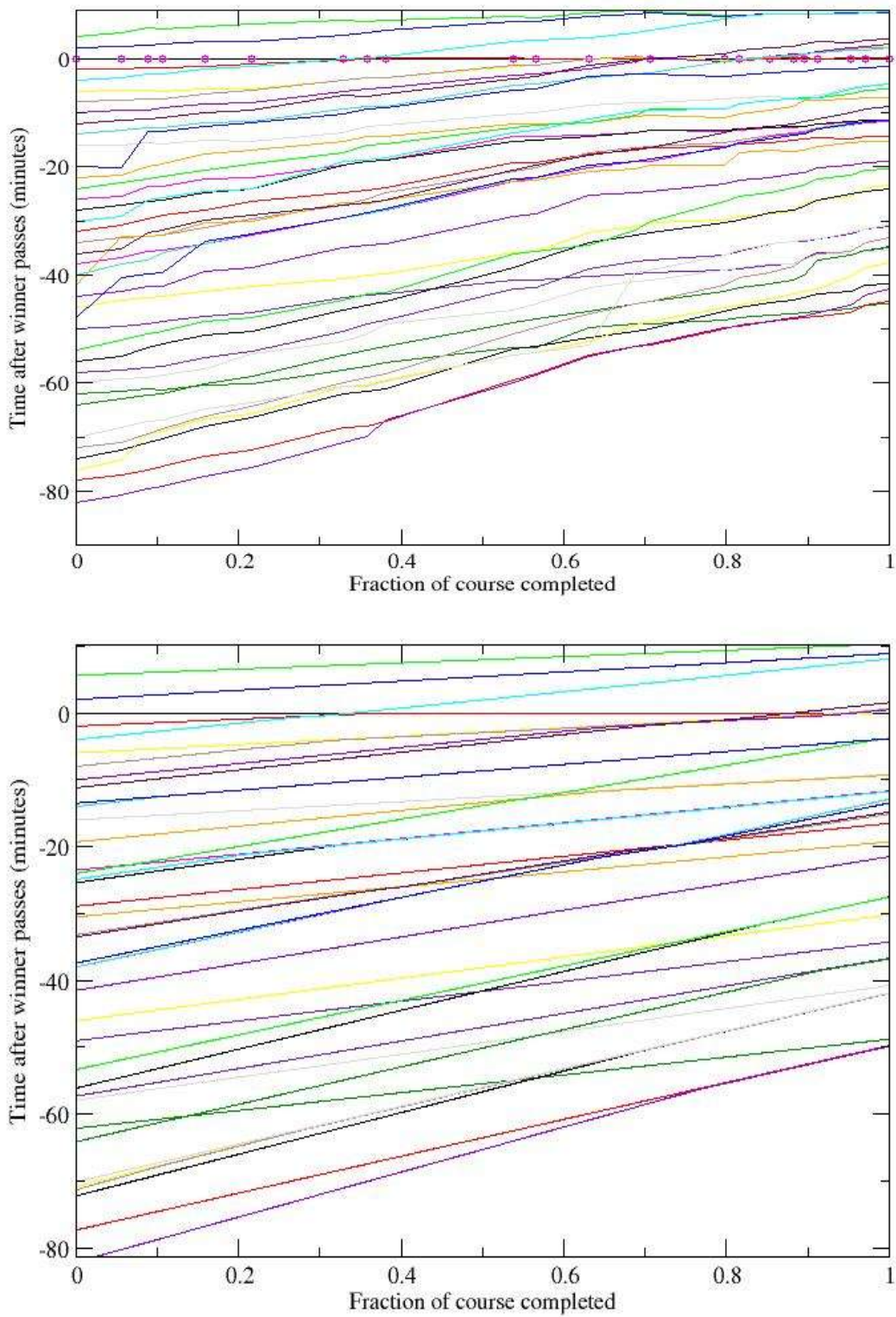

Figure 1. Actual and simulation progression of competitors in the women's classic WOC.
y-axis shows the time relative to the winner at which each competitor is at the location given in the x-axis, hence packs are shown by lines coming together. Colours are consistent between the two figures; competitors named in the text (with start time) are Niggli (black, 0), Jukkola (red, -2), Novikova (blue, -4) Haapakoski (yellow, -6), Allstom (grey, -8), Ryabkina (purple, -10) and Jurenkova (brown, -12). Upper panel shows actual results: circles denote positions of control points where data is recorded. Lower panel shows simulated results, using model described in the text with 8% boost, and minimum time increment of 6sec and range equivalent to 30m. Start times are offset to incorporate errors incurred in the first six legs used to determine $u_i$.

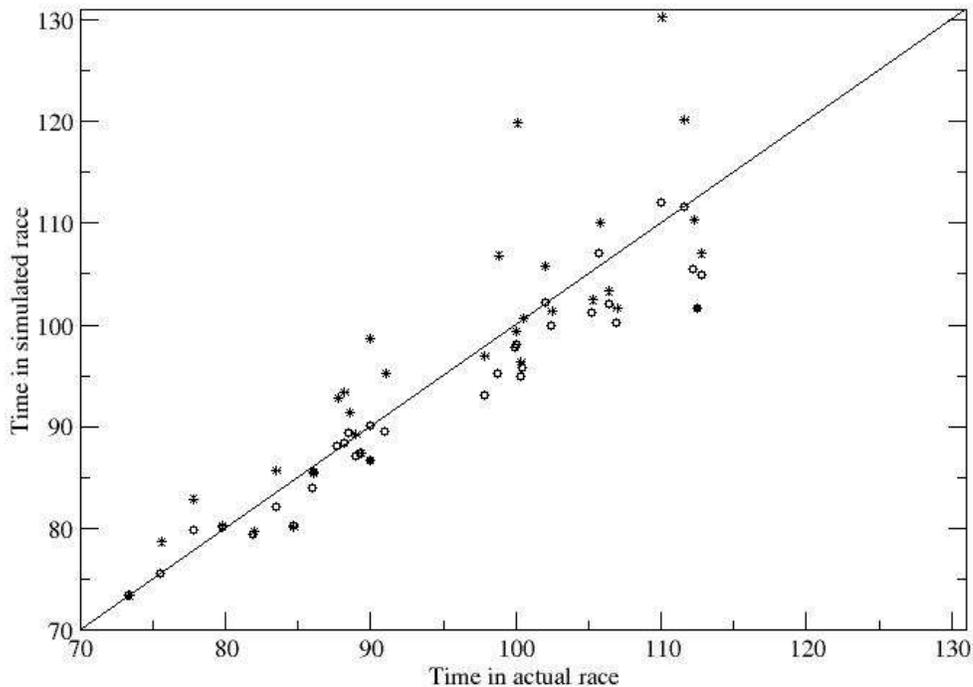

Figure 2. Scatter plot showing correlation between predicted and actual times. Circles are the full models, including packing effects of boost and error suppression, stars represents simple extrapolation from time at the sixth control assuming that independent pace is maintained. Line shows position of perfect agreement. The exact position of control 6 is unknown, so its fractional distance around the course is deduced from the winner's time, which therefore lies exactly on the line. The key point of this figure is that including pack effects gives a significantly better prediction of the final result.

## Discussion

It is surprising that a model based on improved speed from following produces such close results, given the importance of navigational errors in orienteering. However figure 1 reveals a remarkable feature – all errors of more than two minutes are made by orienteers running alone. The men's race has only one such incident. By contrast, most of the competitors spend most of their time in packs.

While the benefit of pack formation in avoiding large errors is not explicitly incorporated in the model, it does have a powerful indirect effect in improving the fit. Since we evaluate our speed data from the first section of the race, errors made then are incorporated in the finishing time. The validity of the approximation that no further errors are made in the latter part of the course is helped by the fact that more packs exist then.

As with all statistical models, the uncertainty of its predictions for specific individuals is such that it is inappropriate to apply them rigorously to individual cases. The actual, simulated and extrapolated women's results all have Niggli and Jukkola as clear leaders. The men's result illustrates how unpredictable events cannot be well predicted by the model: As in the real race, Khramov, Lauenstein and Tavernaro forms early and dominates the high positions, however the winner of the simulated race is, in fact Schneider who forms a pack with Stevenson and maintains his early lead, passing a slower moving pair (Schuler and Horacek). In the actual race these packs did indeed form and meet, however Stevenson suffered an injury. If Stevenson is arbitrarily removed from the model at this point, a Schneider/Schuler pair forms (as in reality) but is not as fast as Schneider/Stevenson and Khramov wins.

Thus, according to the model, Stevenson's injury has far-reaching effects: it eliminates him from the medals, it deprives Schneider of the gold, which goes instead to Khramov, it gains several places for Schuler and costs several places for Horacek.  It is important to emphasize that these specific outcomes are presented as an example, and cannot be deduced with certainty.  However, the more general result is strongly indicated by this work: that a network of strong interactions does exist between competitors, and the actions of one can affect the outcome for numerous others.

Using the model for a race where the distribution of speeds is known it is possible to determine *in advance* whether a particular format with a particular spread of competitor ability through the startlist will be strongly influenced by pack formation.  It can also predict at what stage packing will occur, although not which individuals will benefit.

If the simulation is repeated for the same $u_i$ and starting order, with a 4 minute start interval the onset on pack formation is much delayed: only a small number of pairs are formed late in the race.  While this may appear to resolve the problem of packing, this work also predicts that longer periods of running as individuals would lead to more significant navigation errors, which could increase packing.

 An alternate approach of "forking" the course has been used at previous championships to split pairs of runners from consecutive starts.  Typically this fork has been placed late in the course, however this simulation provides a method for determining the optimal location of the fork, which with a 2 minute start interval is closer to the middle of the course..

## Conclusions

A pre-existing model for pack formation in orienteering has been analysed in the context of the 2005 World Championships.  It has been shown that the formation of packs has a significant outcome on the result.  The effect of pack appears to be twofold, a general increase in speed of around 4-8% (less for the faster members of the pack), and the near elimination of navigational errors involving large time loss.  At most, the benefit derived from the former effect is up to 3-4 minutes, while the benefit from the latter is unquantifiable.